\documentclass[fleqn,usenatbib]{mnras}

\usepackage{newtxtext,newtxmath}

\usepackage[T1]{fontenc}

\DeclareRobustCommand{\VAN}[3]{#2}
\let\VANthebibliography\thebibliography
\def\thebibliography{\DeclareRobustCommand{\VAN}[3]{##3}\VANthebibliography}

\usepackage{graphicx}	
\usepackage{amsmath}	

\usepackage[percent]{overpic}
\usepackage{xcolor}

\newcommand{\chandra}{{\em Chandra}}

\newcommand{\hst}{{\em HST}}
\newcommand{\lum}{erg s$^{-1}$}
\newcommand{\flux}{erg s$^{-1}$ cm$^{-2}$}
\newcommand{\Msun}{M$_{\odot}$}

\title[High-Mass X-ray Binaries and Young Star Clusters]{The Spatial Correlation of High Mass X-ray Binaries and Young Star Clusters in Nearby Star-Forming Galaxies}

\author[B. A. Binder et al.]{
Breanna A. Binder,$^{1}$\thanks{E-mail: babinder@cpp.edu}
Ashley K. Anderson,$^{1}$
Kristen Garofali,$^{2}$
Margaret Lazzarini,$^{3}$ and
\newauthor
Benjamin F. Williams$^{4}$ 
\\
$^1$Department of Physics \& Astronomy, California State Polytechnic University, 3801 W. Temple Ave, Pomona, CA 91768, USA \\
$^2$NASA Goddard Space Flight Center, Code 662, Greenbelt, MD 20771, USA \\
$^3$Cahill Center for Astronomy and Astrophysics, California Institute of Technology, 1200 East California Boulevard, Pasadena, CA 91125, USA \\
$^4$Department of Astronomy, University of Washington, Box 351580, Seattle, WA 98195, USA 
}

\date{Accepted 2023 April 30. Received 2023 March 24; in original form 2023 February 8}

\pubyear{2022}

\begin{document}
\label{firstpage}
\pagerange{\pageref{firstpage}--\pageref{lastpage}}
\maketitle

\begin{abstract}
We present an analysis of the two-point spatial correlation functions of high-mass X-ray binary (HMXB) and young star cluster (YSC) populations in M31 and M33. We find evidence that HMXBs are spatially correlated with YSCs to a higher degree than would be expected from random chance in both galaxies. When supplemented with similar studies in the Milky Way, Small Magellanic Cloud, and NGC~4449, we find that the peak value of the spatial correlation function correlates strongly with the specific star formation rate of the host galaxy. We additionally perform an X-ray stacking analysis of 211 non-X-ray detected YSCs in M31 and 463 YSCs in M33. We do not detect excess X-ray emission at the stacked cluster locations down to 3$\sigma$ upper limits of $\sim10^{33}$ \lum\ (0.35-8 keV) in both galaxies, which strongly suggests that dynamical formation within YSCs is not a major HMXB formation channel. We interpret our results in the context of (1) the recent star formation histories of the galaxies, which may produce differences in the demographics of compact objects powering the HMXBs, and (2) the differences in natal kicks experienced by compact objects during formation, which can eject newly-formed HMXB from their birth clusters. 
\end{abstract}




\section{Introduction}

X-ray binaries (XRBs) are evolved stellar systems composed of a compact object (a neutron star, NS, or a black hole, BH) accreting matter from a companion star. These sources provide a window through which extreme gravity environments can be studied, and the galaxy-wide population properties of XRBs are known to correlate with the host galaxy's total stellar mass \citep[$M_{\star}$;][]{Lehmer+10}, star formation rate \citep[SFR;][]{Ranalli+03, Antoniou+10, Antoniou&Zezas16, Lehmer+19, Mineo+12}, metallicity \citep{Basu-Zych+16,Basu-Zych+13, Brorby+16} and star formation history (SFH) \citep{Boroson+11, Lehmer+19}. XRBs fall into two broad categories, depending on the masses of their companion stars: high-mass XRBs (HMXBs) have massive stellar companions ($M\geq8$\Msun), while those with lower mass companions (typically $\leq2$\Msun) are designated low-mass XRBs (LMXBs). Actively accreting XRBs can achieve X-ray luminosities of $\sim10^{35-39}$ \lum, with the exact luminosity depending on the masses of the two components, the mode of mass transfer occurring within the system, and the accretion rate of material onto the compact object. Ultraluminous X-ray sources (ULXs), with X-ray luminosities in excess of $\sim10^{39}$ \lum, have also been detected in nearby galaxies. XRBs with extremely low accretion rates radiate with X-ray luminosities $\lesssim10^{34}$ \lum\ and are in quiescence, and generally fall below the detection limits of X-ray studies of nearby galaxies. 

HMXBs and LMXBs are generally observed in different environments, and due to the differences in the masses of the donor stars they form and radiate X-rays over very different timescales. Establishing the primary formation channel of HMXBs and LMXBs is therefore of high interest. It is now well-established that LMXB formation is two orders of magnitude more efficient in globular clusters (GCs) than the field \citep{Katz75,Clark75,Fabian+75,Pooley+03}. Due to the high stellar densities in GCs, dynamical formation is the preferred formation channel for LMXBs. By contrast, HMXBs do not appear to preferentially reside within young star clusters (YSCs), but are instead spatially associated with sites of recent and/or ongoing star-forming such as OB associations, YSCs, and H~II regions \citep{Ranalli+03, Kaaret+04, Swartz+04, Lehmer+10, Persic&Rephaeli03, shtykovskiy&gilfanov+07, Walton+11, Mineo+12, Bodaghee+12, Vulic+14, Grimm+03, Bodaghee+21}. The quantity of remaining natal dust and whether or not the young stars are still gravitationally bound to one another are key metrics for distinguishing between YSCs, OB associations, and H~II regions, which can be challenging to measure directly in external galaxies. 

Several scenarios have been proposed to explain the observed spatial correlation of HMXBs with sites of recent star formation: (1) HMXBs receive ``natal kicks'' during the asymmetric supernova explosions that formed the first compact objects, which kicks the systems away from their birthplaces, (2) HMXBs are ejected from their birth clusters by dynamical interactions with other stars in the dense cluster core \citep{mcswain+07}, or (3) HMXBs form within YSCs that eventually disperse. \citet{Bodaghee+12} considered the spatial (or cross-) correlation function between HMXBs and OB associations in the Milky Way, and found that HMXBs were on average $\sim$200-400 pc from the nearest OB association (consistent with the natal kick scenario) with a kinematical age since compact object formation of $\sim$4 Myr. Another recent study by \citet{Fortin+22} identified the birthplaces of $\sim$15 Galactic HMXBs using Gaia (seven of which were YSCs and eight were structures associated with a Galactic spiral arm). However, due to obscuration, only HMXBs out to $\sim$8 kpc can be robustly detected, and astrometric uncertainties and distance uncertainties in the Galactic disk hinder our ability to form a complete census of HMXBs and their orbital trajectories within the Milky Way. 

Nearby star-forming galaxies present opportunities for investigating galaxy-wide HMXB populations and their spatial correlations with YSCs. M31 is the best-studied analog to the Milky Way: its X-ray source population has been observed extensively, with thousands of sources detected and decade-long monitoring campaigns aimed at identifying individual sources and characterizing their population properties \citep{Stiele+11, vulic+13, Vulic+14, vulic+16, DiStefano+04}. Although progress has been made in identifying the nature of individual sources in M31 based on their X-ray properties, their temporal variability \citep{Barnard+14}, or using supervised machine learning techniques \citep{Arnason+20}, the most powerful tool for confidently classifying an X-ray source relies on optical counterpart identification.  

The PHAT and \chandra-PHAT surveys provide the best resource for large-scale studies of HMXBs in M31 \citep[see, e.g.,][]{Lazzarini+21}, while complementary \chandra\ and \hst\ campaigns, including the recent Panchromatic Hubble Andromeda Treasury: Triangulum Extended Region (PHATTER) survey \citep[][Lazzarini et al. {\it submitted}]{Williams+21} and the ChASeM33 survey \citep{Tullmann+11,Garofali+18}, targeted M33. These surveys have resulted in the classification of dozens of XRBs and HMXB candidates in both galaxies. Multi-band photometry was collected for $\sim$100 million stars in M31 by PHAT and for $\sim$22 million stars in M33 by PHATTER, which enabled detailed studies of the galaxies' recent SFHs \citep{Lewis+15,Lazzarini+22}. The citizen science platform Zooniverse\footnote{\url{https://www.zooniverse.org/}} was used to aid in the classification of thousands of star clusters found in the \hst\ imaging in both galaxies \citep{Johnson+12,Johnson+15,Johnson+22,Wainer+22}.

In this work, we analyze the spatial association of HMXBs and YSCs in M31 and M33 and search for X-ray emission from faint or quiescent HMXBs residing within YSCs in these galaxies. In addition to the spatial correlation analysis of HMXBs and OB associations in the Milky Way \citep{Bodaghee+12} and SMC \citep{Bodaghee+21}, \citet{Rangelov+11} presented a systematic study of the HMXB-YSC correlation in the irregular starburst galaxy NGC~4449. These five galaxies together provide a well-studied sample of HMXBs and YSCs that can provide clues to HMXB formation channels and the demographics of compact objects powering the HMXBs. In Section~\ref{section:data} we summarize the relevant properties of the data utilized in this work. In Section~\ref{section:spatial} we describe our spatial correlation analysis between HMXBs and YSC in M31 and M33 and compare our results to other studies of nearby star-forming galaxies. In Section~\ref{section:stacking} we use the technique of stacking to search for faint X-ray emission from YSCs in M31 and M33, and we conclude in Section~\ref{section:conclusions} with a discussion and summary of our findings. 

\section{The Data: High-Mass X-ray Binary and Young Star Cluster Catalogs}\label{section:data}
Our analysis of the HMXB-YSC spatial correlation focuses on M31 and M33, and we supplement our analysis with previously published studies of the Milky Way \citep{Bodaghee+12}, the SMC \citep{Bodaghee+21}, and NGC~4449 \citep{Rangelov+11}. Table~\ref{tab:galaxy_comparison} summarizes the properties of these galaxy, including distance, morphology, optical semi-major axis (characterized by the $D_{25}$ isophotal radius), inclination angle, stellar mass ($M_{\star}$), star formation rate (SFR), and specific SFR (sSFR = SFR/$M_{\star}$). The SFRs reported in Table~\ref{tab:galaxy_comparison} were derived for the last $\sim$100 Myr (i.e., the timescale relevant for HMXBs) using color-magnitude modeling of resolve stellar populations for the SMC, M31, M33 \citep{Rubele+18,Lewis+15,Lazzarini+21}. A combination of FUV and H$\alpha$ emission was used by \citet{Calzetti+18} to estimate the current SFR in NGC~4449, and the reported SFR for the Milky Way represents a meta-analysis of the recent SFR drawn from a large, heterogenous mix of recent SFR indicators \citep{Licquia+15,Chomiuk+11}. The reported SFRs are estimated over approximately the same areas of the galaxies from which we draw our HMXB and YSC samples.

\begin{table*}
    \centering
    \caption{Comparison of Galaxy Properties}
    \begin{tabular}{cccccc}
    \hline\hline
    Property                & Milky Way     & M31       & M33      & NGC~4449 & SMC \\
    (1)                     & (2)           & (3)       & (4)       & (5)       & (6)       \\
    \hline
    Distance                & ...                   & 770 kpc$^a$           & 859 kpc$^b$          & 4.2 Mpc$^c$ & 59.2 kpc$^d$  \\
    Morphology$^e$          & SBbc                  & SA(s)b                & SA(s)cd              & IBm  & SB(s)m pec  \\
    $D_{25}$ (kpc)$^f$   & 22$^g$ & 23 & 8.8 & 3.8 & 2.7 \\
    $M_{\star}$ (\Msun)     & 6$\times10^{10,h}$    & 1$\times10^{11,i}$    & 3$\times10^{9,j}$ & 1$\times10^{9,k}$ & 5$\times10^{8,d}$ \\
    SFR (\Msun\ yr$^{-1}$)  & 1.65$^h$              & 0.3$^l$               & 0.5$^m$               & 0.5$^{n}$ & $0.255^d$ \\
    sSFR$^o$ (yr$^{-1}$)  & $2.75\times10^{-11}$  & $3.00\times10^{-12}$  & $1.67\times10^{-10}$ & $5\times10^{-11}$ &  $5.10\times10^{-10}$  \\
    \hline
    \multicolumn{6}{l}{References: $^a$\citet{McConnachie+05}; $^b$\citet{deGrijs+17}; $^c$\citet{Tully+13};} \\
    \multicolumn{6}{l}{$^d$\citet{Rubele+18}; $^e$Taken from NED; $^f$\citet{RC3};} \\
    \multicolumn{6}{l}{$^g$\citet{Lopez+18}; $^h$\citet{Licquia+15} and \citet{Chomiuk+11}; } \\
    \multicolumn{6}{l}{$^i$\citet{Tamm+12}; $^j$\citet{vanderMarel+12}; $^k$\citet{Querejeta+15};} \\
    \multicolumn{6}{l}{$^l$\citet{Lewis+15}; $^m$\citet{Lazzarini+21}, $^n$\citet{Calzetti+18}; $^o$Computed using the} \\
    \multicolumn{6}{l}{listed values of $M_{\star}$ and SFR} \\
    \end{tabular}\label{tab:galaxy_comparison}
\end{table*}

\subsection{M31}\label{sec:M31_data}
The Panchromatic {\em Hubble} Andromeda Treasury (PHAT) survey \citep{Dalcanton+12} mapped the northern third of M31's star-forming disk using six \hst\ filters from the near-ultraviolet to the near-infrared. It is the largest \hst\ mosaic ever assembled, yielding photometry for over 100 million stars in M31. The Andromeda Project \citep{Johnson+15, Johnson+12} hosted on Zooniverse resulted in a catalog of 2,753 star clusters in M31. In addition to 6-band \hst\ photometry, estimates of cluster ages and masses (derived from modeling the color–magnitude diagrams of the resolved stars in each cluster) are available for roughly half of young (non-globular) PHAT star clusters \citep{Johnson+15,Johnson+17} with sufficiently well-measured photometry in at least three of the \hst\ filters. The complementary \chandra-PHAT survey \citep{Williams+18} used the \chandra\ ACIS-I detector to image nearly the entire PHAT footprint in the X-rays down to a limiting 0.35-8 keV flux of $\sim3\times10^{-15}$ \flux\ (corresponding to a luminosity of $\sim3\times10^{35}$ \lum\ at the distance of M31). The \chandra\ X-ray sources were aligned to optical sources in the PHAT catalog to a positional precision of better than $\sim$0.1$^{\prime\prime}$, which corresponds to a distance of $\sim$0.37 pc at the distance of M31. Using the combined power of \hst, \chandra, and {\em NuSTAR}, \citet{Lazzarini+18,Lazzarini+21} identified 58 HMXBs within the PHAT survey footprint according to their X-ray properties, the optical and UV properties of their optical counterparts, and their association with sites of recent star formation. Of the 373 X-ray sources detected by the \chandra-PHAT survey, 185 sources lacked an optical counterpart. Based on scaling relations \citep[e.g., ][]{Lehmer+14}, \citet{Williams+18} estimate $\sim$100 of these sources to be low-mass XRBs. The identity of the remaining $\sim$85 sources is unknown, although the majority are likely highly extincted background AGN based on statistical estimates of the AGN number densities \citep{Luo+17}. It is therefore unlikely that a significant number of HMXB candidates remain unidentified in the area covered by the PHAT and \chandra-PHAT surveys.

\begin{figure}
     \centering
     \includegraphics[width=0.9\linewidth]{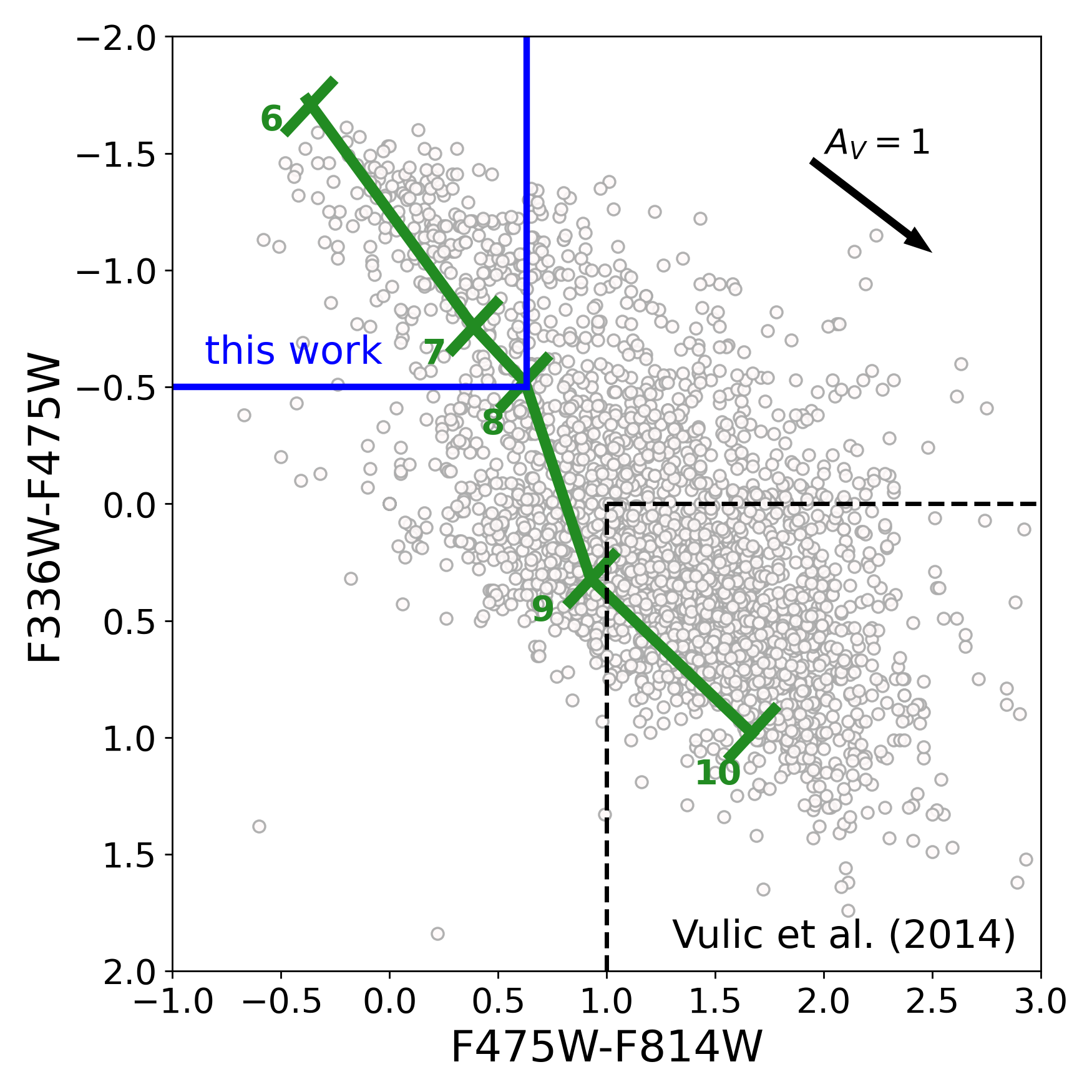}
     \caption{Color-color diagram of PHAT star clusters from \citet{Johnson+15}. A theoretical evolutionary track from a 10$^4$ \Msun\ stellar population using PARSEC \citep{Bressan+12} and COLIBRI \citep{Marigo+13} is shown in green (numbers indicate log(age)), assuming $Z=0.02$ and a total reddening of $E(B-V)=0.13$ \citep{Caldwell+11}. The YSC sample used in this work is located within the blue box in the upper-left corner of the diagram, while the dashed box in the lower-right corner shows the region of the diagram containing the 83 star clusters considered in \citet{Vulic+14}. The foreground reddening vector is of length $A_V=1$ mag.}
     \label{fig:cluster_properties}
\end{figure}

\begin{figure*}
    \centering
\begin{tabular}{cc}
    \includegraphics[width=0.48\linewidth,clip=true,trim=2cm 6cm 2cm 5cm]{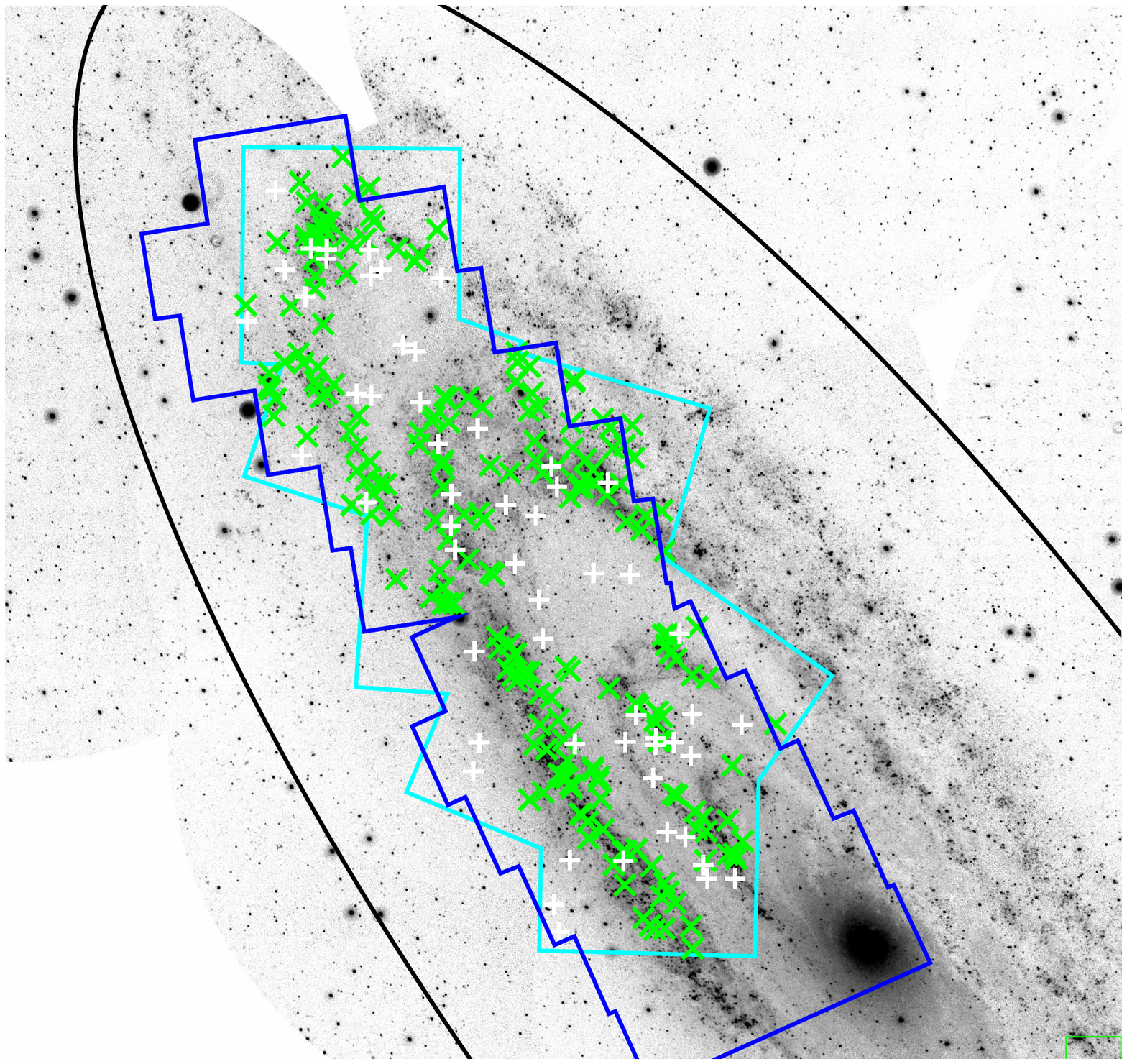} &
    \includegraphics[width=0.48\linewidth,clip=true,trim=2cm 6cm 2cm 5cm]{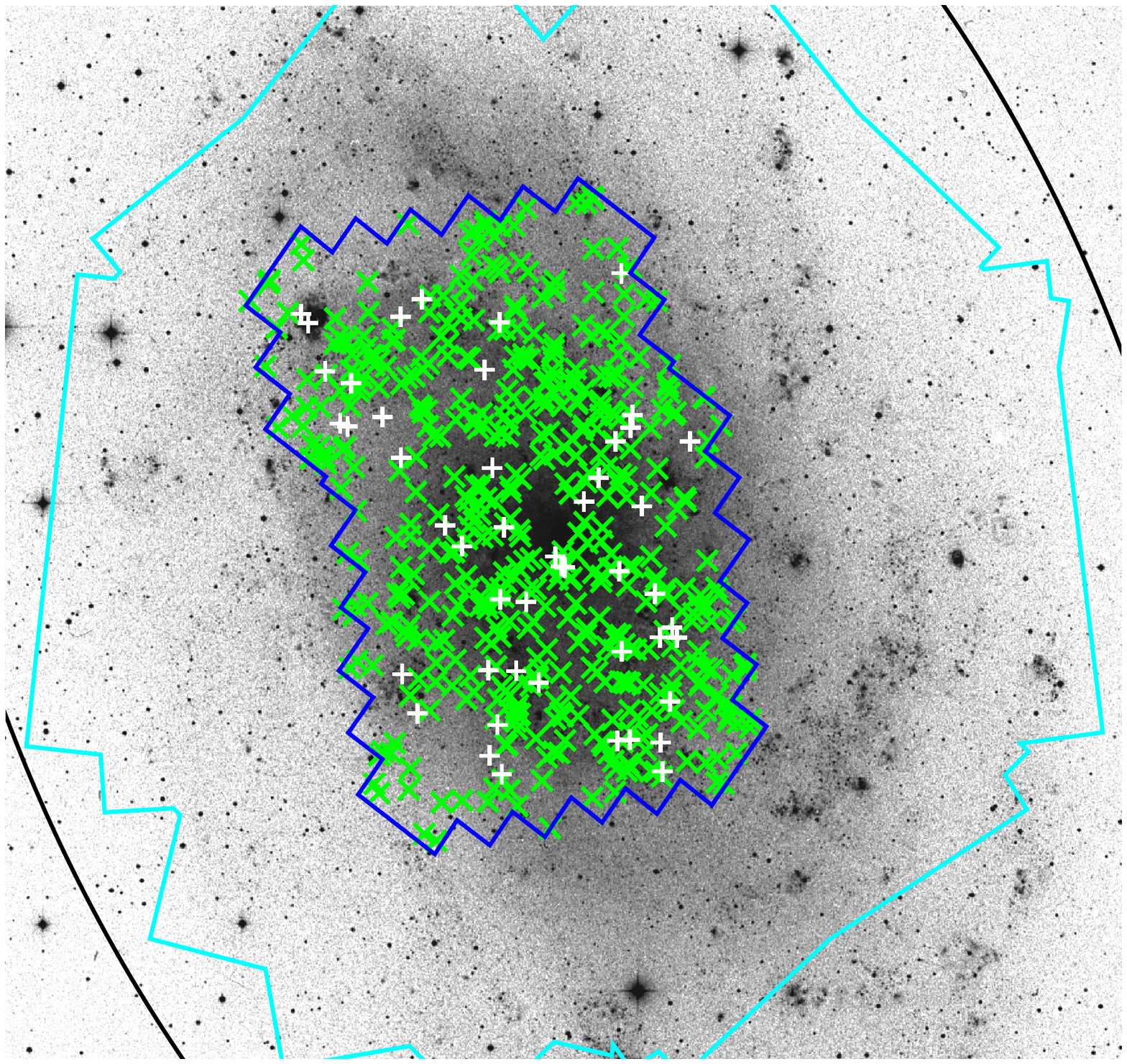} \\
\end{tabular}
    \caption{Comparison of YSC and HMXB locations in M31 (superimposed on the GALEX NUV image; {\it left}) and M33 (superimposed on a DSS image; {\it right}). White crosses indicate the locations of HMXBs and green Xs indicate the locations of the YSCs in both panels. The PHAT and PHATTER survey footprints are shown in blue in M31 and M33, respectively, and the \chandra\ footprint for the \chandra-PHAT survey ({\it left}) and ChASeM33 survey ({\it right}) are shown in cyan. The black outlines indicate the $D_{25}$ isophote for each galaxy.}
    \label{fig:YSC_HMXB_map}
\end{figure*}

The PHAT cluster catalog contains a mixture of both young ($\lesssim$10 Myr) and older ($\gtrsim$1 Gyr) star clusters, as estimated from colors and comparison to synthetic YSCs \citep{Johnson+15}. The association of XRB and older star cluster connection was previously studied by \citet{Vulic+14}; in this work, we aim to restrict our analysis to a systematically younger sample that is more likely to share an evolutionary history with HMXBs. The massive stars that form both the compact objects and donor stars in HMXBs have main sequence lifetimes $\lesssim$100 Myr, so star clusters older than this are not associated with current HMXB populations. To define our YSC sample, we require that a cluster fall entirely within the \chandra-PHAT footprint and be detected in the \hst\ filters F336W, F475W, and F814W. We follow the same procedure as \citet{Vulic+14} to generate theoretical evolutionary tracks for a $10^4$ \Msun\ star cluster \citep[the approximate cluster mass upper limit for clusters with ages 10-100 Myr in the PHAT cluster catalog;][]{Johnson+17} using PARSEC \citep{Bressan+12} and COLIBRI \citep{Marigo+13}, assuming a metallicity of $Z=0.02$ (appropriate for M31) and a total reddening (both internal and external) of $E(B-V)=0.13$ \citep{Caldwell+11}. Figure~\ref{fig:cluster_properties} shows a color-color diagram of the PHAT star clusters with this evolutionary track overlaid. Star clusters $\lesssim$100 Myr typically have F475W-F814W$\leq$0.63 and F336W-F475W$\leq$-0.5 \citep[shown by the blue lines in the figure;][]{Johnson+15}. We therefore require that clusters fall within this bluest portion of this color-color diagram. The resulting YSC sample contains 258 clusters that are significantly bluer, brighter, and likely much younger in age than the clusters considered in \citet{Vulic+14}. Figure~\ref{fig:YSC_HMXB_map} shows the locations of the HMXBs (in white) and YSCs (in green) that we use in our analysis.

\subsection{M33}
M33 was observed extensively in X-rays with the \chandra\ ACIS Survey of M33 \citep[ChASeM33;][]{Tullmann+11,Plucinsky+08}, which imaged $\sim$70\% of the galaxy's $D_{25}$ isophote with a total exposure time of 1.4 Ms. Over 660 X-ray sources were detected down to a limiting 0.35-8 keV luminosity of $\sim2\times10^{34}$ \lum. The PHATTER survey covered nearly the entire M33 disk out to $\sim1.5-2$ scale lengths, providing 6-band \hst\ imaging for $\sim$22 million stars \citep{Williams+21}. From this combined \chandra\ and \hst\ imaging, Lazzarini et al. {\it (submitted)} identified 62 HMXBs and strong HMXB candidates \citep[see also][]{Garofali+18} in M33. The positional alignment between the two surveys is similar to that of M31 ($\sim0.1^{\prime\prime}$, corresponding to a distance of $\sim$0.42 pc at the distance of M33). Although the Lazzarini et al. ({\it submitted}) HMXB catalog represents a subset of all the HMXB candidates identified to-date with \hst, using sources identified only from the PHATTER survey \citep[rather than archival \hst\ coverage, i.e.,][]{Garofali+18} ensures highly precise astrometric alignment between HMXB candidates and YSCs, more uniform \hst\ coverage (in terms of both depth and filters used), and makes a better direct comparison sample to M31. In this work, we use only the HMXB candidates with optical spectral energy distributions consistent with that of massive star, and we require that the sources do not raise any of the ``flags'' that may indicate an alternate, non-HMXB origin for the X-ray emission (such as having extremely soft X-ray hardness ratios, having mismatched values of $A_V$ and $N_H$ inferred from optical and X-ray observations, or having an optical counterpart with a spectral type that is not consistent with being a massive star). Eight X-ray sources were found coincident with optical sources in the PHATTER data but raised multiple flags -- these sources are HMXB candidates, but additional observations are required to determine the identity of these sources and we do not utilize these sources in our analysis. As with the PHAT imaging of M31, the Local Group Cluster Search citizen science initiative used \hst\ imaging obtained by the PHATTER survey to construct a catalog of 1,216 star clusters in M33 \citep{Johnson+22}. CMD modeling of 729 star clusters in M33 imaged in at least three \hst\ filters yielded cluster ages of $6.08<$log(age/yr)$<8.91$, with a median log(age/yr) value of 7.96. In our analysis, we use only the 464 clusters with best-fit ages $<$100 Myr. Again, the M33 HMXBs (white) and YSCs (green) used in our analysis are shown in Figure~\ref{fig:YSC_HMXB_map}.

\section{Spatial Correlation of High-Mass X-ray Binaries and Young Star Clusters}\label{section:spatial}
\subsection{Nearest Neighbor YSC to an HMXB}
\begin{figure*}
    \centering
\begin{tabular}{cc}
     \includegraphics[width=0.43\linewidth,clip=true,trim=0cm 0.3cm 0.3cm 0cm]{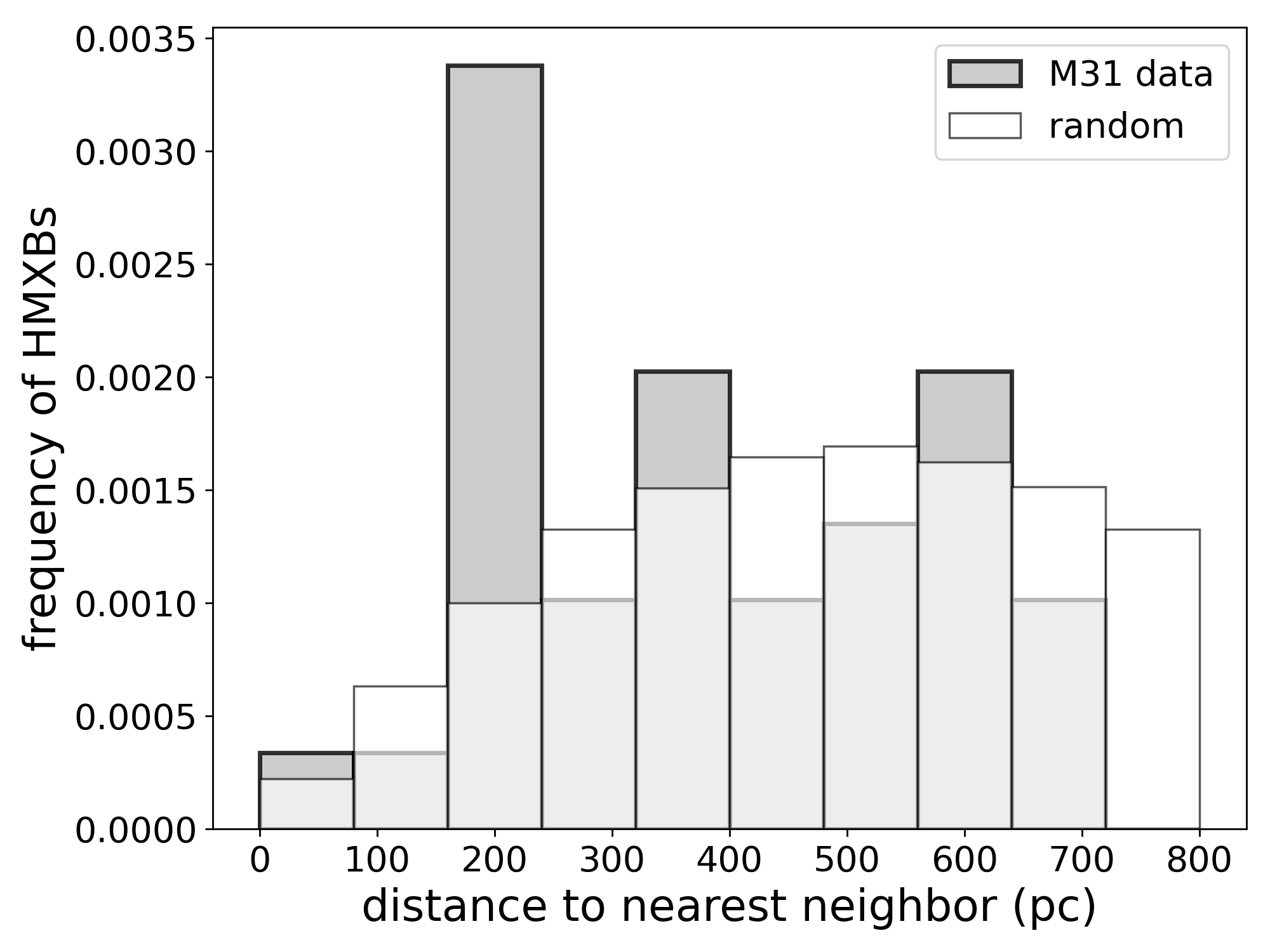} &
     \includegraphics[width=0.43\linewidth,clip=true,trim=0.1cm 0.3cm 0.3cm 0cm]{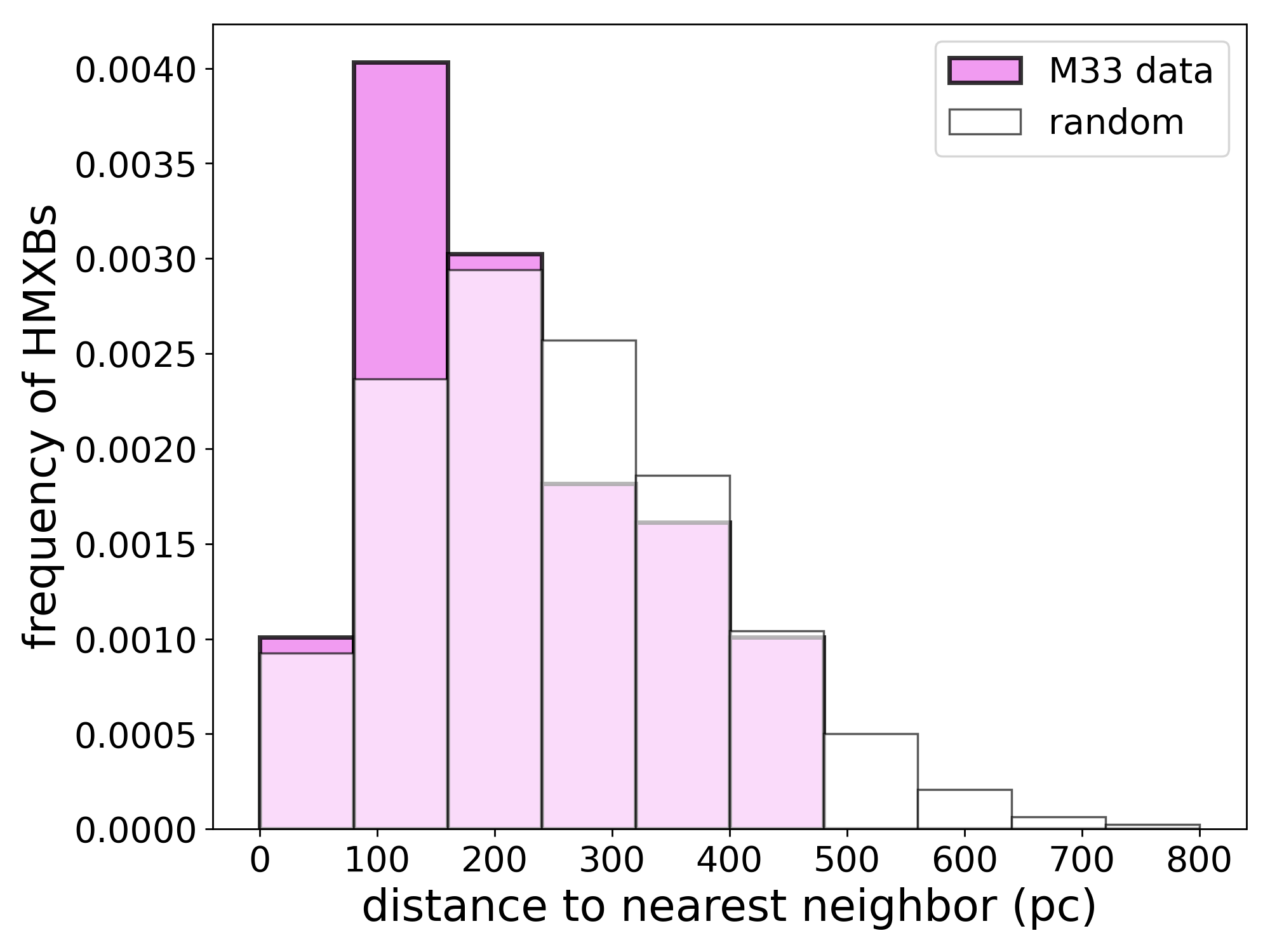}  \\
\end{tabular}
    \caption{Distribution of distances between HMXBs and their nearest YSC neighbors for M31 ({\it left}, gray) and M33 ({\it right}, pink). Distributions of distances compared to randomly, homogeneously-distributed YSC locations within the survey area is shown are white.}
    \label{fig:nearest_neighbor}
\end{figure*}

To begin our analysis of the spatial distributions of HMXBs and YSCs, we first identify the nearest neighboring YSC to each HMXB and measure the projected and inclination-corrected separation distance (hereafter referred to as ``data-data,'' or DD, pairings). Figure~\ref{fig:nearest_neighbor} shows a histogram of the nearest neighbor distance between the HMXBs and YSCs for both M31 and M33. The DD distributions are shown in gray and pink for M31 and M33, respectively. We additionally generate homogeneously-distributed, random YSC locations within the respective footprint of each galaxy, and calculate the nearest neighbor random YSC separation distance (which we refer to as ``data-random,'' or DR, pairings). This process is performed 5,000 times, and the resulting average DR distribution is shown in white in both panels. We additionally measure the mean separation between HMXBs and YSCs (both observed and randomly distributed) in each galaxy.

A summary of statistics comparing the DD and DR distributions for the minimum and mean separation distances between HMXBs and YSCs is presented in Table~\ref{tab:clustering_results}. We additionally provide the observed (0.35-8 keV) X-ray luminosity range of the HMXBs used in our analysis. As seen in Figure~\ref{fig:nearest_neighbor}, there is an excess in the fraction of HMXBs that are separated from their nearest observed YSC neighbor by $\lesssim$200 pc compared to the randomized sample pairings. In both galaxies, the DD pairings show smaller mean separations than the DR pairings, although the spread in nearest neighbor distances are large for both samples. We also compute higher order moments in the nearest neighbor separation distributions (e.g., the skew and kurtosis) to quantify potential differences in the overall shapes of the DD an DR distribution. Skewness measures deviations from a perfectly symmetrical distribution (with positive values indicating a greater number of smaller values than expected from a normal distribution), while kurtosis measures whether a distribution is too peaked (positive values indicate a distribution that is more peaked than expected from a normal distribution). In M31 in particular, both the minimum and mean separation distributions of the real data have kurtosis values significantly less than that of the DR distributions generated by random, homogeneously distributed YSC samples. The same trend is present in the mean separation distance distributions in M33 -- the DD distribution is less peaked than the DR distribution -- but the minimum separation distance distributions are much more comparable (and overall closer to normal distributions) than in M31.

\begin{table*}
    \centering
    \caption{HMXB-YSC Minimum and Mean Separation Distance Statistics in M31 and M33}
    \begin{tabular}{cccccccccccc}
    \hline\hline
            &        &      & log$L_X^{\rm HMXB}$      && \multicolumn{3}{c}{Minimum Separation Distance (pc)} && \multicolumn{3}{c}{Mean Separation Distance (kpc)}  \\  \cline{6-8} \cline{10-12}
    Galaxy & \# YSC & \# HMXBs & [\lum]    && mean  & skew  & kurtosis      && mean & skew  & kurtosis  \\
    (1)    & (2)     & (3)      & (4)   && (5)   & (6)   & (7)   && (8)       & (9)  & (10)        \\
    \hline
    M31     & 258   & 58  & 35.5-37.0   & DD & 0.58$^{+0.48}_{-0.35}$  & 1.84 & 3.75 && 6.78$\pm$1.16 & 1.34 & 3.03 \\
            &       &     &   & DR & 0.60$^{+0.37}_{-0.30}$  & 2.82 & 13.16 && 9.53$\pm$0.96 & 1.83 & 6.29 \\
    \hline
    M33     & 464   & 62  & 34.8-37.0  & DD & 0.19$^{+0.15}_{-0.08}$   & 0.82 & -0.13 && 5.25$^{+1.69}_{-0.47}$ & 1.55 & 1.79  \\
            &       &     &   & DR & 0.24$^{+0.15}_{-0.12}$  & 0.63 & 0.26 && 6.87$^{+1.50}_{-0.22}$ & 1.79 & 3.05 \\
    \hline
    \end{tabular}\label{tab:clustering_results}
\end{table*}

\subsection{The Spatial Correlation Function}
\begin{figure}
    \centering
    \includegraphics[width=1\linewidth,clip=True,trim=0cm 2cm 0cm 1cm]{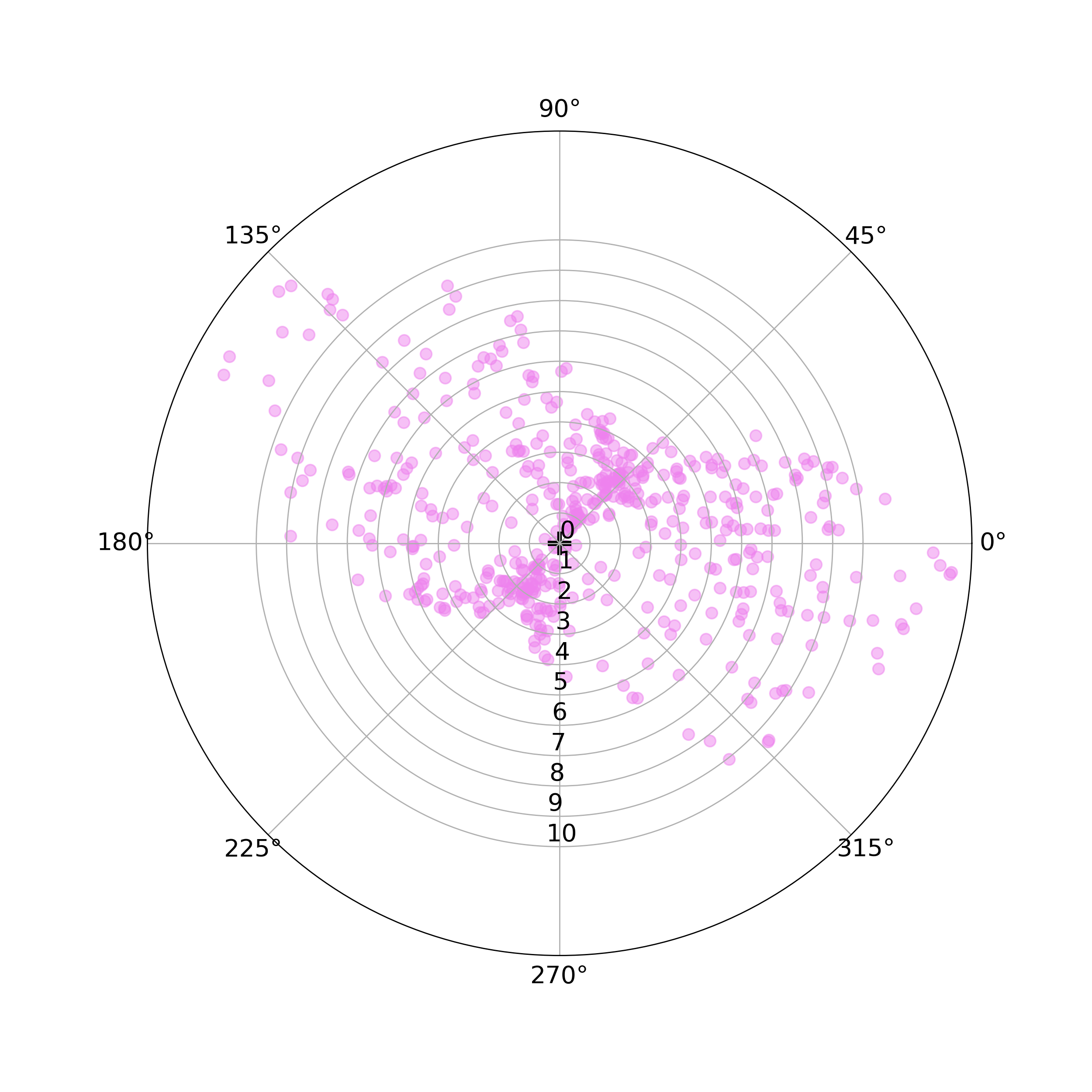}
    \caption{An example of constructing $N_{DD}$ for an HMXB in M33 (black cross at the center). The pink circles show the deprojected locations of YSCs relative to the HMXB, with concentric circular rings shown (in units of kpc). We have only labeled integer radial bins for clarity; our analysis (described in the text) uses rings of 0.5 kpc thickness. The choice of $0^{\circ}$ is arbitrary.}
    \label{fig:example_radial}
\end{figure}

We next computed the spatial correlation function to evaluate the degree of clustering between HMXBs and YSCs in M31 and M33, following the same methodology described in \citet{Bodaghee+12,Bodaghee+21}. For each HMXB, we construct concentric rings centered on each HMXB of 0.5 kpc thickness out to 10 kpc and count the number of observed YSCs found in each ring. We tested several ring thicknesses (100 pc, 250 pc, 0.5 kpc, and 1 kpc) and found this did not qualitatively affect our results, although the uncertainties were larger for thinner rings containing relatively few HMXB-YSC pairings; we therefore used the 0.5 kpc thickness so our results could be easily compared to the most recent similar study in the SMC \citep{Bodaghee+21}. Figure~\ref{fig:example_radial} shows an example of this process, with one of the M33 HMXBs at the center and the distribution of deprojected YSC distances in radial bins (we only label integer bins for clarity, and the 0$^{\circ}$ is arbitrary). These represent the number of ``data-data'' pairings ($N_{DD}$) in this section. We carry out the same procedure using the true location of each HMXB and a mock YSC population which we refer to as the ``data-random'' pairings, $N_{DR}$, and a randomized HMXB location and the observed YSC population (the ``random-data'' pairings, $N_{RD}$). We finally use the randomized YSC and HMXB samples generated in the previous steps to compute ``random-random'' pairings, $N_{RR}$. To compute the spatial correlation $\xi(r)$ between HMXBs and YSCs, where $r$ represents the distance of each 0.5 kpc thick radial bin from its central HMXB, we use the \citet{Landy&Szalay93} definition:

\begin{equation}
    \xi(r) = \frac{N_{DD} - N_{DR} - N_{RD} + N_{RR}}{N_{RR}},
\end{equation}

\noindent which has the advantage of nearly Poissonian variance. The function $\xi(r)$ provides a quantitative measurement of the degree of spatial clustering of YSCs about the positions of the HMXBs. For calculations involving random HMXBs and YSCs, we use the same number of random sources as real sources in our sample. This processes is repeated 5,000 times, with each iteration having different randomized populations, and report the mean spatial correlation functions and the 90\% confidence intervals. We note that this process generates a total of $\sim1.3\times10^6$ mock YSCs and $\sim2.9\times10^5$ mock HMXBs in M31, and $\sim2.3\times10^6$ mock YSCs and $\sim3.1\times10^5$ mock HMXBs in M33 with which we compare the observed data.

Randomized YSC catalogs can be constructed in a number of ways. For both M31 and M33, we use three different approaches to construct mock YSC catalogs. First, we define a homogeneous sample in which YSCs are randomly distributed across the PHAT or PHATTER footprints with a uniform surface density. We next use the scale lengths of M31 \citep[$\sim$5.3 kpc;][]{Dalcanton+12} and M33 \citep[$\sim$1.5 kpc;][]{Regan+94} and the central coordinates of each galaxy to construct an exponentially decaying profile that approximately follows the light distribution of the stellar disks. We distribute YSCs according to this exponentially decaying profile (excluding the nuclear region), so that the number density of YSCs is higher closer to the center of the galaxy and falls off with increasing galactocentric radius (this is hereafter referred to as the exponential sample). Finally, we construct a ``bootstrap'' sample, where the mock catalog is constructed by randomly selecting right ascension and declination values from the observed YSC catalogs to form new coordinate pairs (in other words, the right ascension of one YSC is randomly paired with the declination value of a different cluster). Bootstrap resampling is frequently employed by extragalactic spatial correlation studies \citep[e.g.,][]{Gilli+05, Meneux+09, Krumpe+10}. Mock HMXBs are randomly distributed with a uniform surface density across the observed area. In all randomized catalogs, mock sources may not fall outside of the observed area of either galaxy.

\begin{figure}
\centering
\includegraphics[width=1\linewidth]{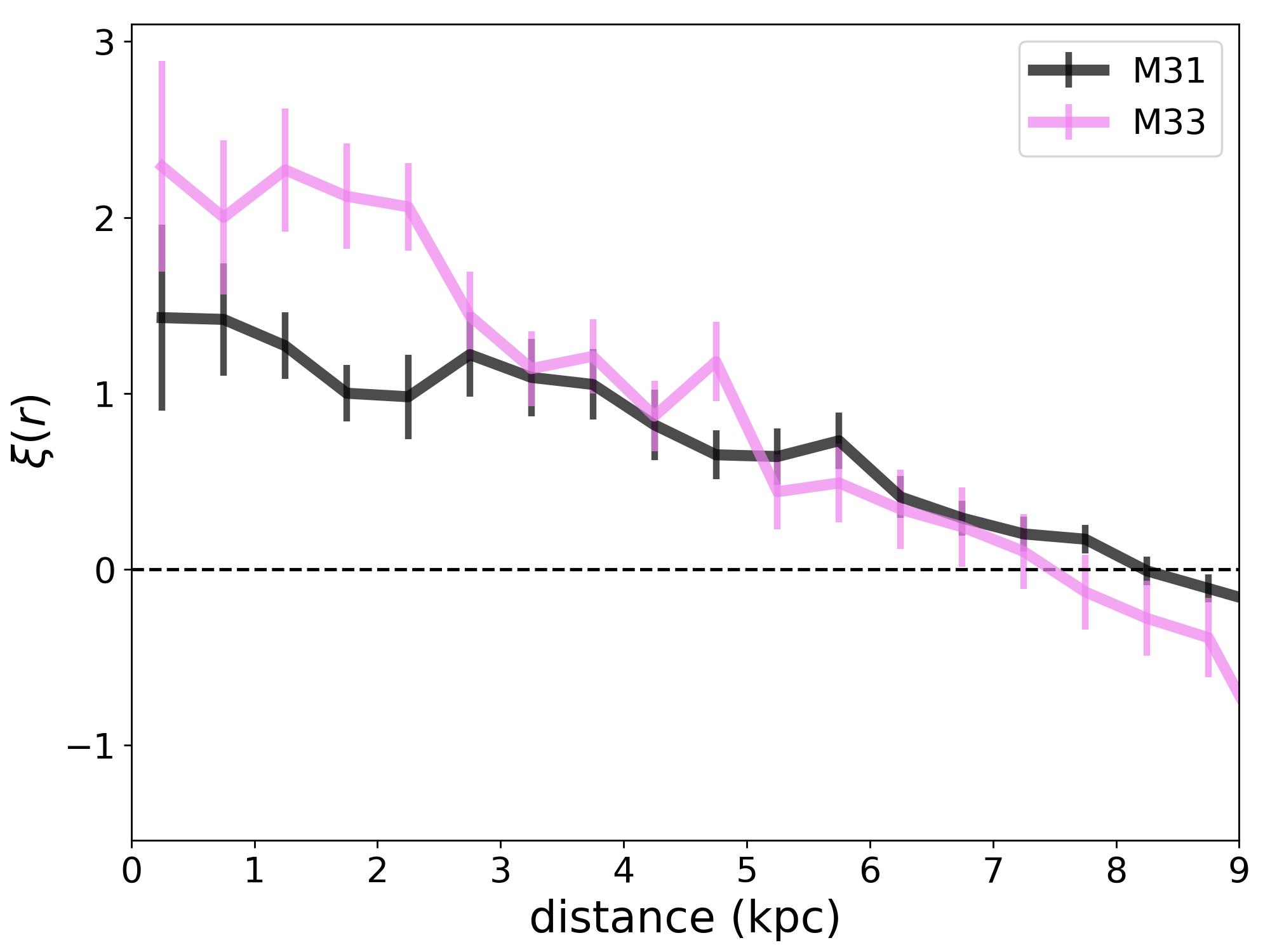}
\caption{The spatial correlation function for M31 (black) and M33 (pink), showing a clear excess of YSCs located at distances $\lesssim$6-7 kpc from an HMXB. The horizontal dashed line at $\xi=0$ represents the value expected for no spatial correlation.}
\label{fig:HMXB_excess}
\end{figure}

We compute spatial correlation functions for both galaxies using each of the three mock YSC catalogs. The spatial correlation functions derived from the homogeneous mock YSC catalogs for both galaxies are shown in Figure~\ref{fig:HMXB_excess}. In both galaxies, the probability of finding a YSC near an HMXB is significantly higher than expected from Poisson statistics. In the innermost bin ($r<0.5$ kpc), the observed clustering $\xi$ between HMXBs and YSCs is $\sim4\sigma$ above the value expected for no correlation ($\xi=0$) in M33 and $\sim2.8\sigma$ above $\xi=0$ in M31. The enhancement in YSCs near HMXBs exceeds 3$\sigma$ significance for distances less than $\sim$6 kpc and $\sim$5 kpc in M31 and M33, respectively, consistent with the results found for the Milky Way \citep{Bodaghee+12} and SMC \citep{Bodaghee+21}. The spatial correlation functions derived using the exponential and bootstrap mock catalogs did not show evidence of significant clustering, suggesting that these mock catalogs provided much better descriptions of the overall YSC distributions than the homogeneously distributed YSC sample. We further tested introducing ``extraneous'' X-ray sources into the DD sample, by inserting the locations of X-ray sources that are unlikely to be HMXBs into our list of HMXBs. This effect dilutes the clustering signal: we find that the peak value of $\xi(r)$ decreases by $\sim$0.12 (about 20\% the size of the uncertainties) for every non-HMXB source introduced into the DD pairing.

\subsection{Results}
The measured values of $\xi(r<0.5$ kpc) are $1.43\pm0.53$ and $2.29\pm0.66$ for M31 and M33, respectively. This implies that one counts approximately two and a half times as many observed YSCs as randomized ones in M31 and just over three times as many YSCs near HMXBs in M33. This is comparable to the results obtained for the Milky Way \citep{Bodaghee+12} and the SMC \citep{Bodaghee+21}. The cumulative distribution of the displacement between HMXBs and YSCs in NGC~4449 \citep{Rangelov+11} also exhibits a similar trend: there are roughly 3-5 times as many observed HMXBs and YSCs within several hundred parsecs of each other than expected from randomly distributed sources. It is becoming increasingly apparent that the locations of HMXBs and YSCs in star-forming galaxies are highly correlated, with typical separations of a few hundred pc. We note that this is likely a lower estimate of the average separation distance between an HMXB and its parent YSC, as the nearest YSC may not be the actual parent cluster to a given HMXB. The HMXB may not have formed in a YSC at all; YSCs are typically surrounded by larger star-forming complexes and/or OB associations which could be the true birth site of the HMXBs, as was found for some Galactic HMXBs \citep{Fortin+22}.

Does the degree of spatial correlation between HMXBs and YSCs correlate with host galaxy properties? If the HMXBs experience a ``kick'' during and asymmetric supernova during compact object formation, then the separation between an HMXB and its birth cluster increases with time since compact object formation. We would therefore expect younger (measured relative to the time of compact object formation) HMXBs to be located systematically closer to their birth clusters than older HMXBs. A galaxy experiencing a high SFR will have recently produced more young stars (and, therefore, YSCs and HMXBs) than a galaxy with a lower recent SFR; the degree of very close clustering between HMXBs and YSCs may therefore correlate with the recent SFR of the host galaxy. Observations of large samples of star-forming galaxies show a tight correlation between SFR and galaxy stellar mass $M_{\star}$ \citep[e.g., see][and references therein]{Ilbert+15}. In order to examine the connection between star formation activity and the distribution of HMXBs within the galaxy, we compute the sSFR using the SFR and $M_{\star}$ values reported in Table~\ref{tab:galaxy_comparison}. The five galaxies summarized in Table~\ref{tab:galaxy_comparison} have sSFRs spanning from $\sim3\times10^{-12}$ yr$^{-1}$ in M31 to $\sim5\times10^{-10}$ yr$^{-1}$ in NGC~4449. We note that the SFRs and stellar masses assumed here were not measured over the same observed area of the galaxies as $\xi(r)$. However, none of these galaxies have undergone significant recent mergers or exhibit pronounced asymmetries in their star-forming disks, so the sSFR is unlikely to vary significantly in different subgalactic regions within the galaxy.

We adopt a value of $\xi(r<1$ kpc)$=1.61\pm0.60$ for the Milky Way \citep{Bodaghee+12} and 
$\xi(r<0.5$ kpc)$=5.15\pm0.60$ for the SMC \citep[][which uses the same functional form of $\xi$ and the same homogeneously distributed random clusters and OB associations as used in this work]{Bodaghee+21}, and approximate $\xi(r<0.5$ kpc)$=4\pm1$ for NGC~4449 from the analysis in \citet{Rangelov+11}. \citet{Bodaghee+12} did not report $\xi(r<0.5 {\rm kpc})$ for the Milky Way, as many of the HMXBs considered in that work had distance uncertainties larger than 0.5 kpc. If we assume the $\xi(r)$ function in the Milky Way follows the same general shape as in the SMC, M31, and M33, we would expect $\xi(r<0.5 {\rm kpc})$ to be higher than the $\xi(r<1 {\rm kpc})$ value, although likely still within the reported uncertainties. We therefore do not expect this difference in the $\xi$ measurements of the Milky Way would significantly impact our results.

\begin{figure}
    \centering
    \includegraphics[width=1\linewidth]{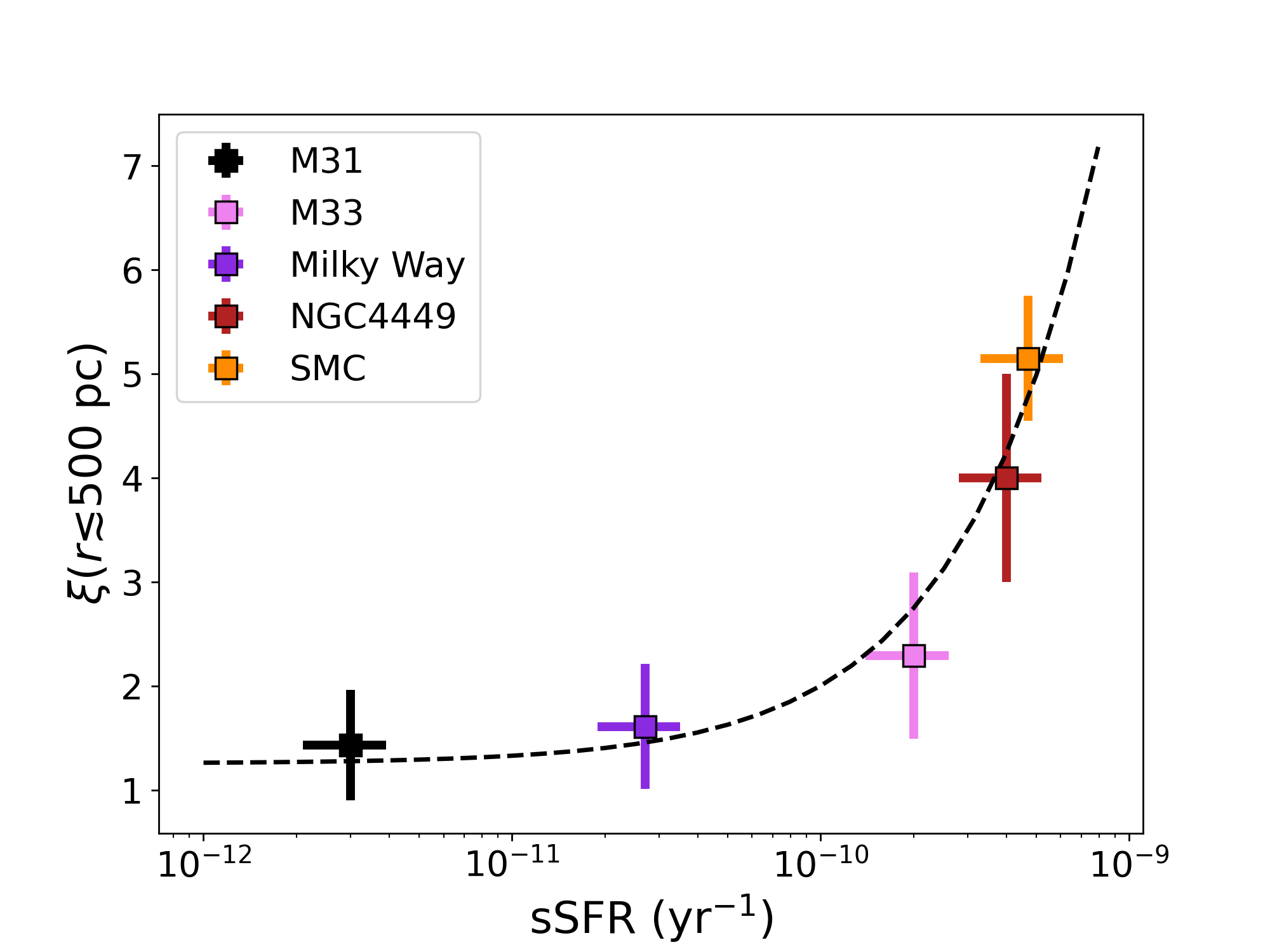}
    \caption{The peak value of the spatial correlation function $\xi$ between HMXBs and YSCs as a function of sSFR (computed using the values reported in Table~\ref{tab:galaxy_comparison}) of the host galaxy. Values for the Milky Way \citep{Bodaghee+12}, NGC~4449 \citep{Rangelov+11}, and the SMC \citep{Bodaghee+21} are taken from the literature. The best-fit linear model is shown by a dashed line.}
    \label{fig:xi_sSFR}
\end{figure}

In Figure~\ref{fig:xi_sSFR} we plot the peak value of $\xi$ as a function of sSFR. There is a clear correlation between the two parameters: increasing the sSFR results in a significantly stronger spatial correlation between the YSCs and HMXBs in the galaxy. To characterize this relationship, we fit a simple linear model to $\xi$ as a function of sSFR. The best fit relationship is given by

\begin{equation}\label{eq:xi_sSFR}
    \xi = \left(0.75\pm0.09\right) \frac{\text{sSFR}}{10^{-10}~\text{yr}^{-1}} + \left(1.25\pm0.27\right).
\end{equation}

\noindent This relationship predicts the excess probability (i.e., the probability in excess of Poisson) of finding an HMXB within $\sim$0.5 kpc of a YSC in a galaxy, given the sSFR of the galaxy. One important caveat to this relationship is that it not generally applicable to extremely low-sSFR galaxies; in the limit of zero recent star formation, the intercept in the above equation predicts $\xi>0$, which would be indicative of spatial clustering. The sSFR of a galaxy must be high enough for both YSC and HMXB production to occur. In an analysis of HMXB X-ray luminosity functions in a sample of 29 nearby star-forming galaxies (containing $\sim$700 XRBs), \citet{Mineo+12} found that the number of bright ($L_X>10^{35}$ erg s$^{-1}$) HMXBs in a galaxy depends on the galaxy's SFR as $\approx 135\times$SFR. This relationship predicts $\sim$40-68 HMXBs residing in M31 and M33, consistent with the observed number, and suggests that the SFR of a galaxy must be greater than $\sim0.007$ \Msun\ yr$^{-1}$ to form any HMXBs at all. LMXBs are expected to be the dominant population in galaxies with sSFR $\lesssim10^{-12}$ \citep{Lehmer+19}; we therefore only expect the above relationship to hold for galaxies with sSFRs above this threshold.

We note that there is a metallicity dependence implicit in Figure~\ref{fig:xi_sSFR} (and in Equation~\ref{eq:xi_sSFR}). Given the galaxy mass-metallicity relationship, the lowest-mass galaxies in our sample will also have the lowest metallicities. Although HMXB production efficiency has been shown to depend on metallicity \citep{Douna+15}, the metallicity dependence is strongest for only the very brightest sources -- i.e., lower metallicity environments are observed to host a larger number of ULXs with $L_X>10^{39}$ erg s$^{-1}$. None of the galaxies considered in this work host such luminous sources, and the X-ray luminosity function of HMXBs at lower luminosities ($\sim10^{36-38}$ erg s$^{-1}$, which encompasses most of the HMXBs considered here) is not strongly metallicity-dependent \citep{Lehmer+21}.

\section{Searching for Quiescent X-ray Binaries Residing in Young Star Clusters}\label{section:stacking}
Although there is a clear correlation between the locations of HMXBs and YSCs in star forming galaxies, relatively few bright HMXBs are observed to reside {\it within} YSCs \citep[the notable exception to this in our sample is NGC~4449, the galaxy with one of the highest sSFRs;][]{Rangelov+11}, and there is no evidence that YSCs with higher masses or higher stellar densities produce more HMXBs than lower-mass, lower-stellar density clusters of the same age \citep{Mulia+19}. Only one YSC in M33 \citep[cluster \#29 from the PHATTER cluster catalog, which has an age of $\sim$10 Myr;][]{Johnson+22, Wainer+22} is found to be coincident with an X-ray source \citep[which was detected by {\it XMM-Newton}, but is not contained in the ChASeM33 catalog;][]{Pietsch+04,Misanovic+06}.

The relative lack of HMXBs residing in their birth clusters could be due to age, the ratio of BH-powered to NS-powered HMXBs, metallicity effects, or a combination of these three factors. In order for an HMXB to appear spatially coincident with a YSC, it must be old enough for compact object formation to have occurred, but young enough to not have traveled far from its birth cluster. BHs may receive smaller (or zero) kicks during their formation compared to NSs, and thus BH-HMXBs are more more likely than NS-HMXBs to remain associated with their natal clusters. Low metallicity environments are found to be correlated with higher HMXB production rates in general \citep{Douna+15}. NGC 4449 has all the ingredients to satisfy high likelihood of detecting HMXBs coincident with YSCs: it is a relatively low metallicity galaxy (comparable to the LMC) that experienced a strong burst of star formation $\sim$10 Myr ago \citep{Sacchi+18}, and hence hosts a large population of very young YSCs and young HMXBs in which BHs primaries are likely over-represented \citep{Rangelov+11}. 

While bright HMXBs ($L_X \gtrsim10^{35}$ \lum) may only rarely be found coincident with a YSC, the question of whether quiescent HMXBs (with $L_X\lesssim10^{34}$ erg s$^{-1}$) reside in their birth clusters is relatively open. \citet{Vulic+14} analyzed over 1 Ms of \chandra\ observations of M31 and performed an X-ray stacking analysis of star clusters and H~II regions to search for faint, quiescent XRBs residing within their parent clusters, which resulted in non-detections down to a limiting X-ray luminosity of $\sim10^{32}$ \lum. However, the star clusters considered in \citet{Vulic+14} are systematically older than the YSCs considered in this work, and no such stacking analysis has yet been performed using ChASeM33 observations of M33. 

We therefore performed an X-ray stacking analysis of the YSCs in M31 and M33, utilizing \chandra\ observations from the \chandra-PHAT and ChASeM33 projects. A full discussion of the data reduction process for each survey is provided in \citet[for M31]{Williams+18} and \citet[][for M33]{Tullmann+11}; we briefly summarize the relevant data processing procedures of each study and discuss the stacking techniques we use to search for faint HMXBs within YSCs in these galaxies. All observations were reduced using standard procedures with CIAO \citep[using version 4.7 for M31 and 4.0.1 in M33;][]{Fruscione+06}. Exposure maps and images were generated using \texttt{fluximage}, and point-spread function maps were made using \texttt{mkpsfmap}. For both M31 and M33, the \chandra\ images were aligned to the optical \hst\ imaging data, resulting in greatly improved astrometry \citep[a precision of better than 0.1$^{\prime\prime}$;][Lazzarini et al. {\it submitted}]{Williams+18}. The limiting 0.35-8 keV flux of each individual ACIS-I pointing in the \chandra-PHAT survey is $\sim3\times10^{-15}$ \flux, which corresponds to a luminosity of $\sim3\times10^{35}$ \lum\ at the distance of M31; the limiting 0.35-8 keV luminosity of the ChASeM33 survey was $\sim2\times10^{34}$ erg s$^{-1}$ \citep{Tullmann+11}.

\begin{table}
\setlength{\tabcolsep}{3pt}
    \centering
    \caption{Summary of \chandra\ Observations Used in Stacking Analysis}
    \begin{tabular}{cccccc}
    \hline \hline
      & R.A.    & Decl.   &  & Exp. Time &  \\
    ObsID  & (J2000) & (J2000) & Date & (ks) & \# YSCs \\
    (1)     & (2)       & (3)   & (4)   & (5)  & (6) \\
    \hline
    \multicolumn{6}{c}{{\bf M31}: \chandra-PHAT Observations from \citet{Williams+18}} \\
    \hline
    17008 &  00:44:15.70 & 41:23:15.3 & 2015 Oct 06    & 49.1 & 51 \\ 
    17009 &  00:44:04.17 & 41:34:39.0 & 2015 Oct 26    & 49.4 & 20 \\ 
    17010 &  00:44:59.06 & 41:32:03.6 & 2015 Oct 19    & 49.4 & 19 \\ 
    17011 &  00:45:30.16 & 41:43:24.7 & 2015 Oct 08    & 49.4 & 24 \\ 
    17012 &  00:44:46.98 & 41:51:39.1 & 2015 Oct 11    & 48.4 & 33 \\ 
    17013 &  00:46:08.28 & 41:57:28.6 & 2015 Oct 17    & 44.8 & 32 \\ 
    17014 &  00:46:21.13 & 42:09:16.9 & 2015 Oct 09    & 49.1 & 32 \\ 
    \hline
    \multicolumn{4}{r}{Total Exp. Time (Ms):}   & 10.22  & 211 \\ 
    \hline \hline
    \multicolumn{6}{c}{{\bf M33}: ChASeM33 Observations from \citet{Tullmann+11}} \\
    \hline
    6376 & 01:33:51.14 & +30:39:20.5 & 2006 Mar 03 & 94.3 & 355 \\ 
    6377 & 01:33:50.18 & +30:39:51.3 & 2006 Sep 25 & 93.2 & 58 \\ 
    6378 & 01:34:13.21 & +30:48:02.9 & 2005 Sep 21 & 95.5 & 32 \\ 
    6382 & 01:33:08.20 & +30:40:10.6 & 2005 Nov 23 & 72.7 & 4 \\ 
    6385 & 01:33:27.40 & +30:31:40.6 & 2006 Sep 18 & 90.4 & 6 \\ 
    6386 & 01:34:06.49 & +30:30:26.7 & 2005 Oct 31 & 14.8 & 4 \\ 
    7402 & 01:34:13.47 & +30:48:04.2 & 2006 Sep 07 & 45.2 & 4  \\ 
    \hline
    \multicolumn{4}{r}{Total Exp. Time (Ms):}   & 43.01 & 463 \\
    \hline\hline
    \end{tabular}
    \label{tab:stacking_summary}
\end{table}

\begin{figure*}
    \centering
    \begin{tabular}{cc}
        \includegraphics[width=0.48\linewidth]{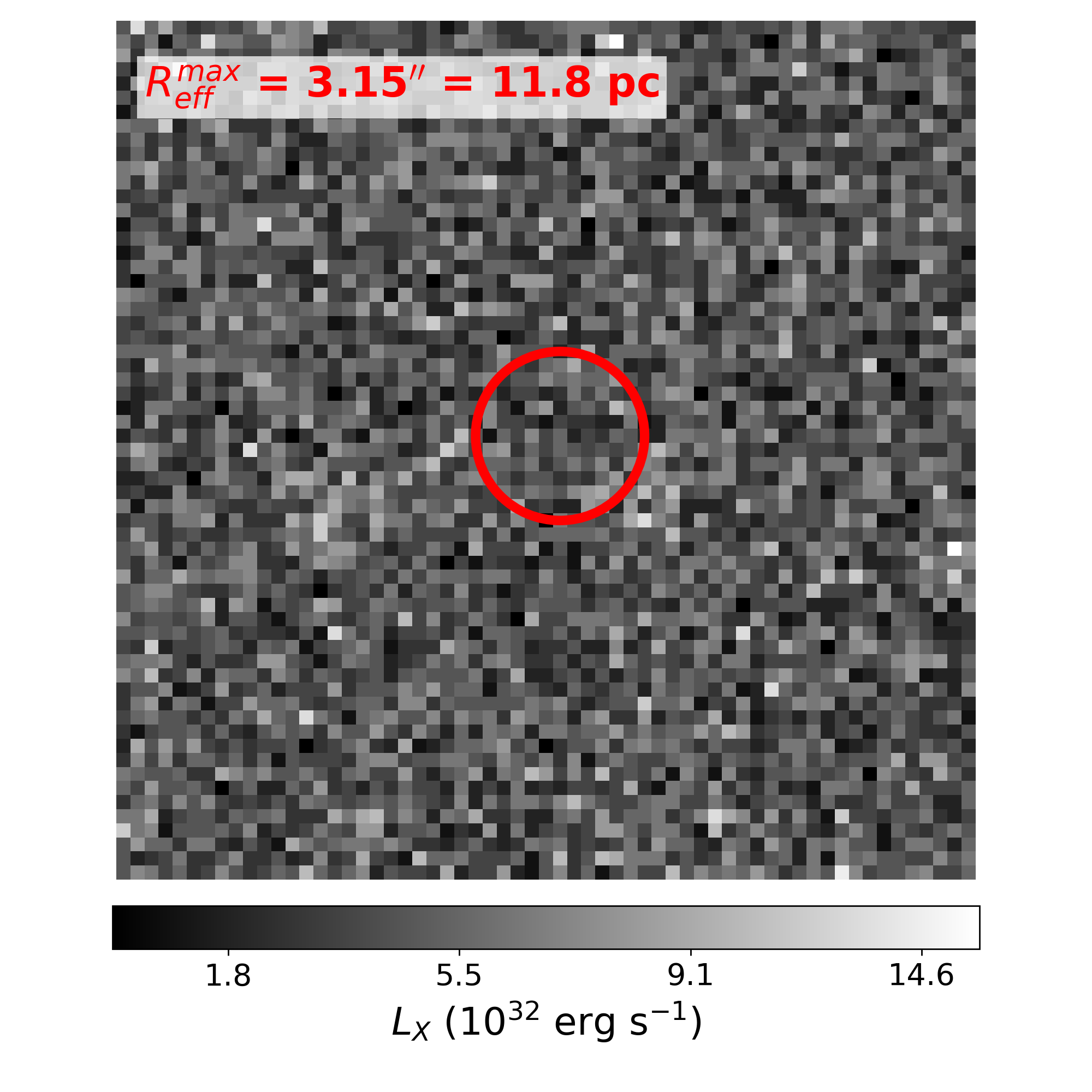} & \includegraphics[width=0.48\linewidth]{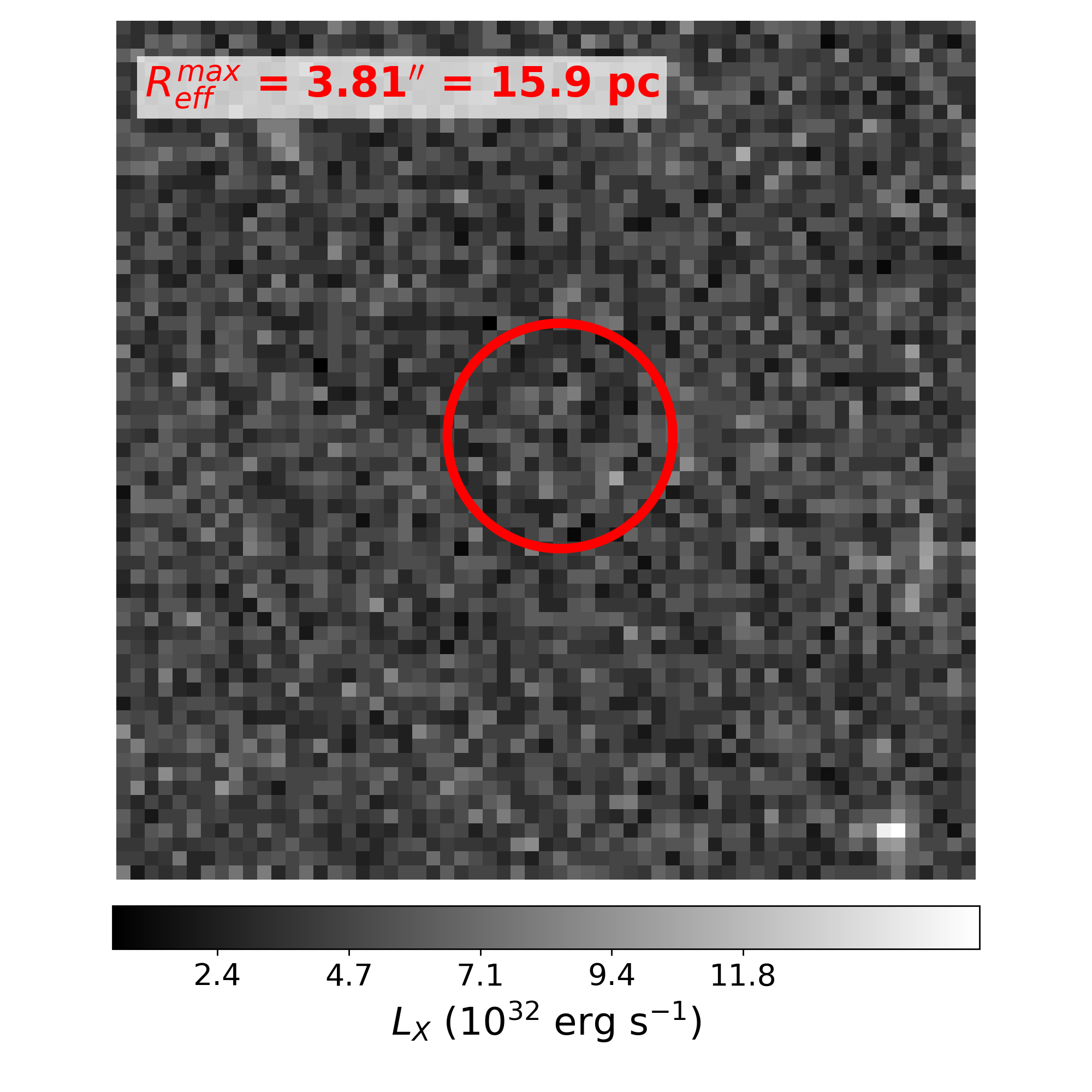} \\
        \includegraphics[width=0.48\linewidth]{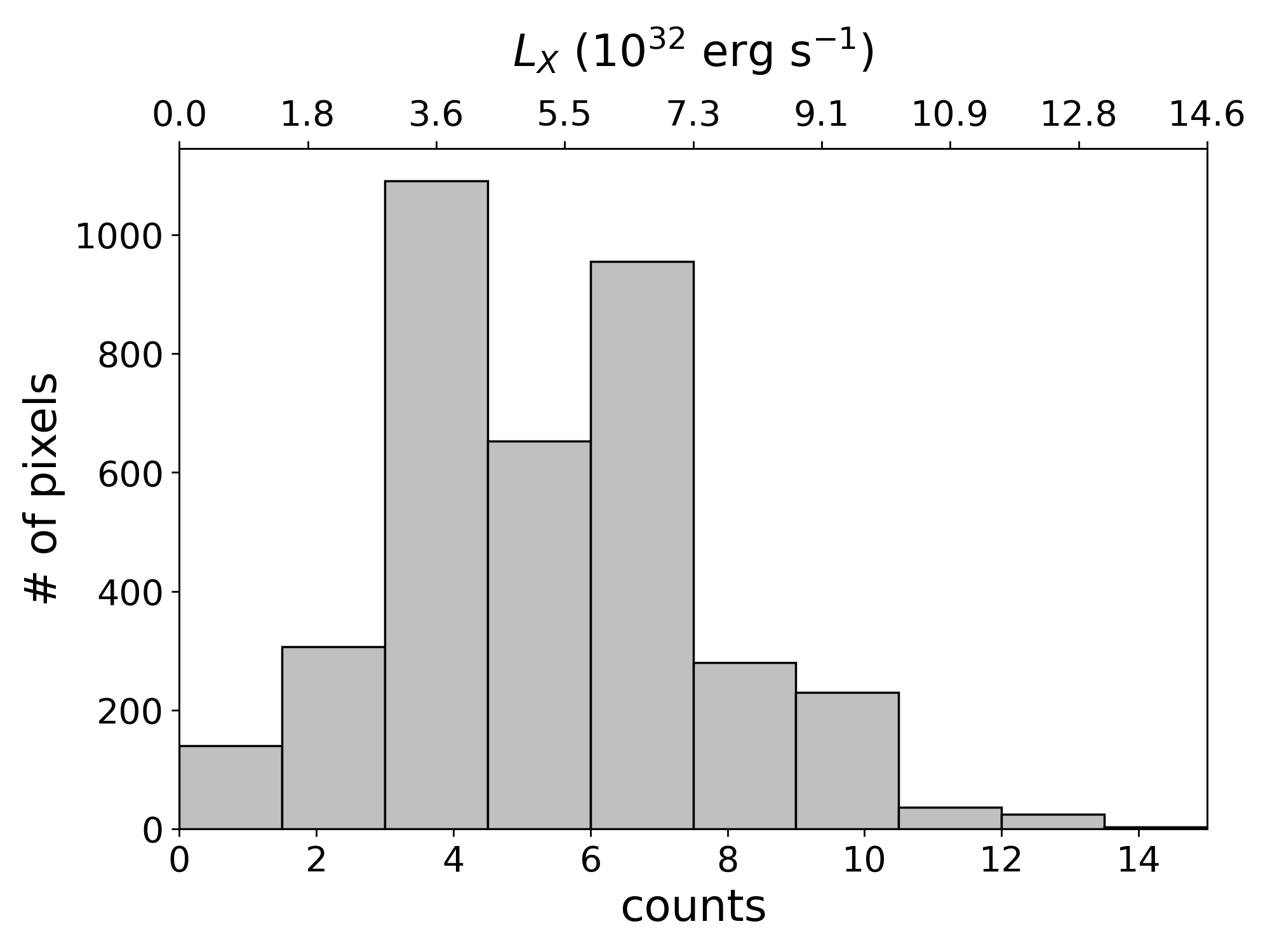} & \includegraphics[width=0.48\linewidth]{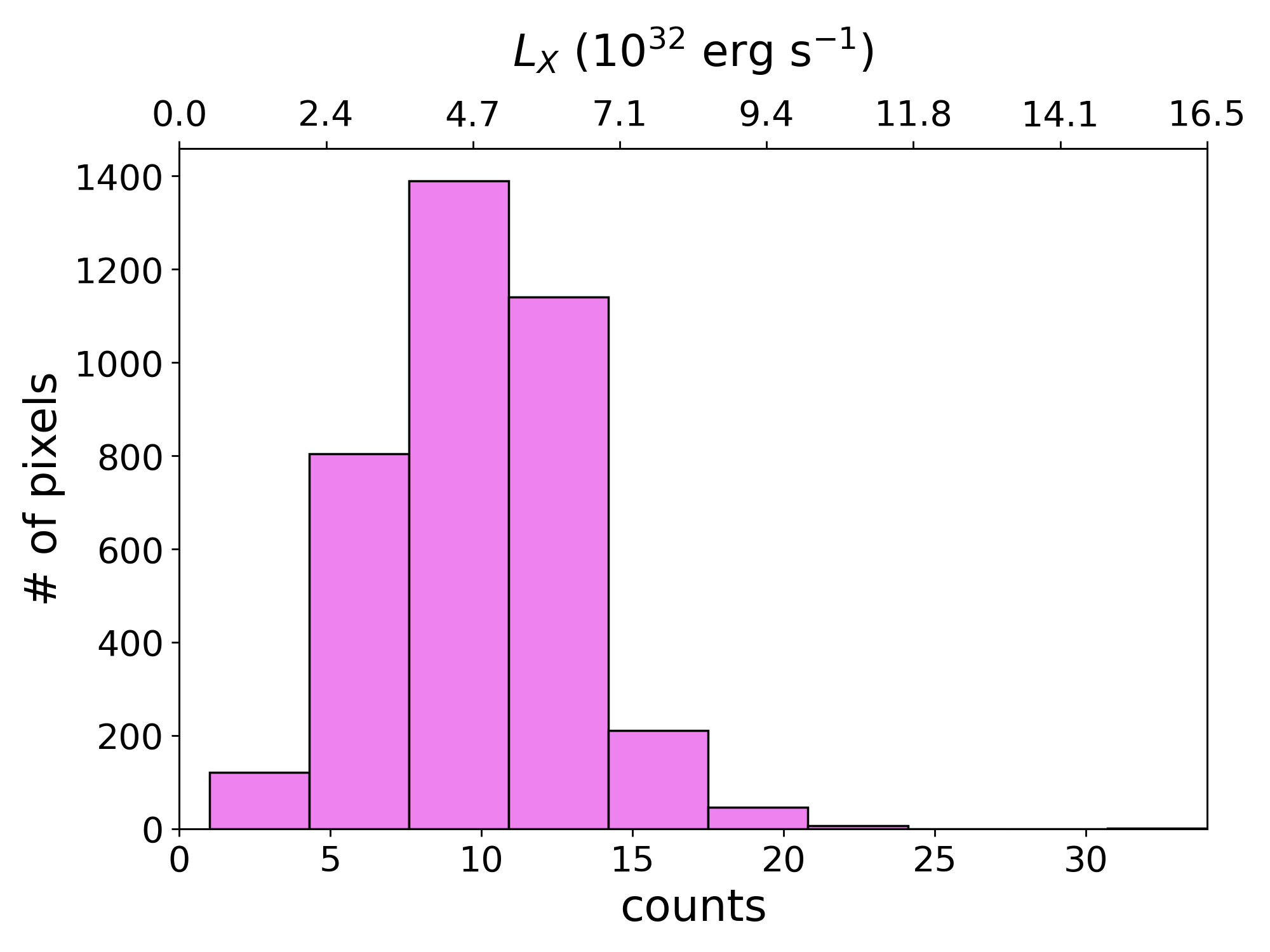} \\
    \end{tabular}
    \caption{{\it Top}: Stacked \chandra\ images of 211 YSCs in M31 ({\it left}) and 463 YSCs in M33 ({\it right}). The red circles corresponds to the maximum effective radii of a YSC in each sample \citep{Johnson+15,Johnson+22}. {\it Bottom}: Histograms showing the distribution of counts in the stacked images for M31 ({\it left}, gray) and M33 ({\it right}, pink).}
    \label{fig:stack_all}
\end{figure*}

Of the 258 YSCs in M31 used in the spatial correlation analysis from Section~\ref{section:spatial}, 211 ($\sim82$\%) fall within the \chandra-PHAT footprint (for a total effective exposure time of 10.22 Ms). None of the X-ray imaged clusters was found to be coincident with an X-ray source. Relevant information about the \chandra-PHAT fields containing at least one YSC is summarized in Table~\ref{tab:stacking_summary}. To construct a stacked X-ray image, we first created postage stamp \chandra\ images (63$\times$63 pixels) of each star cluster, centered on the cluster location given in \citet{Johnson+15}. We then used the CIAO task \texttt{dmregrid2} to stack all individual postage stamps into a final merged image. The resulting stacked image is shown in Figure~\ref{fig:stack_all}, along with a histogram showing the distribution of counts measured in the resulting image; no obvious X-ray point sources are detected at the stacked cluster location (shown in red). To convert the image from count rate to a 0.35-8 keV energy flux, we use the same conversion factor as in \citet[][count rate$\times1.313\times10^{-11}$]{Williams+18}. This energy flux is then be converted to an X-ray luminosity using the distance to M31. The final stacked image shows an average X-ray luminosity of $(4.7\pm2.1)\times10^{32}$ \lum. The 3$\sigma$ upper limit on the average X-ray luminosity at the location of the merged star clusters in M31 is therefore $\sim1.1\times10^{33}$ \lum. 

Following the same approach as for M31, we create postage stamp \chandra\ images of each star cluster location in M33 and use \texttt{dmregrid2} to combine all postage stamp images into a final merged image. Of the 464 YSCs identified in Section~\ref{section:spatial}, only one YSC was {\it not} observed in the ChASeM33 footprint; the total effective exposure time for the 463 X-ray imaged YSCs is 43.01 Ms. We use an energy conversion factor of count rate$\times2.295\times10^{-11}$ to convert the image into units of 0.35-8 keV luminosity \citep[assuming a power law spectrum with $\Gamma=1.7$, and a Galactic absorbing column of 1.85$\times10^{21}$ cm$^{-2}$;][]{HI4PI+16}. Again, the final stacked image and histogram of observed counts are shown in Figure~\ref{fig:stack_all}. The stacked image shows no evidence of excess X-ray emission at the stacked cluster location: the average X-ray luminosity in the image is (4.6$\pm$1.5)$\times10^{32}$ \lum. The 3$\sigma$ upper limit on the X-ray luminosity is $\sim9\times10^{32}$ \lum. 

To obtain even deeper effective exposure times, we additionally perform a stacking analysis of all non-X-ray detected star clusters in M31 and M33. We exclude known globular clusters and all clusters within $\sim4^{\prime\prime}$ of an X-ray source so as not to artificially raise the inferred background X-ray luminosity in the final stacked image. We use 2,197 clusters (total effective exposure time of $\sim$106 Ms) in M31 and 1,203 clusters (effective exposure time $\sim$112 Ms) in M33 and follow the same procedure described above. The resulting stacked images show no evidence for point-like X-ray emission at the center of the stacked star clusters. The 3$\sigma$ upper limits on the X-ray luminosities are $\sim7\times10^{32}$ erg s$^{-1}$ in M31 and $\sim9\times10^{32}$ erg s$^{-1}$ in M33. 

Despite the significantly deeper effective exposure times, the full-star cluster sample 3$\sigma$ upper limits on the X-ray luminosities are comparable to those found for the YSC samples only, indicating that we may be detecting the diffuse, low-luminosity X-ray background due to stars and star formation in M31 and M33. The dominant source of this X-ray emission will evolve over time: for example, stellar winds and supernovae in YSCs will result in soft, diffuse emission with X-ray luminosities on the order of $\sim10^{32-33}$ erg s$^{-1}$ on timescales $\lesssim$40 Myr \citep{Cervino+02,Oskinova+05,Townsley+11}. The most massive O-type and Wolf-Rayet stars will themselves be sources of X-ray emission, albeit for only a few million years. The lack of an X-ray detection at the stacked cluster location in the YSC sample suggests that there are relatively few {\it very} young ($\lesssim$10 Myr) and very massive YSCs in the sample, where we may expect to observe a brighter diffuse X-ray component due to winds and/or supernovae. It is therefore likely that the X-ray upper limits of the stacked YSC samples are dominated by X-ray emission from massive stars within the clusters. The X-ray upper limits on the full cluster stacked sample, which contains many more older star clusters, is likely X-ray emission from a much wider range of stellar masses.

\section{Discussion and Conclusions}\label{section:conclusions}
Our analysis of the HMXB and YSC populations in nearby galaxies shows that (1) HMXBs and YSCs are spatially correlated with one another, and the peak value of the spatial correlation function correlates with the sSFR of the host galaxy; and (2) there is no evidence for a population of quiescent HMXBs residing within YSCs down to a limiting luminosity of $\sim10^{33}$ \lum\ (0.35-8 keV) in both M31 and M33. The lack of an X-ray detection in the stacked YSC images suggests that the number of HMXBs still residing in their parent YSCs is minimal in these galaxies, and that dynamical formation  within YSCs is not a major HMXB formation channel \citep[consistent with][]{Mulia+19}. HMXBs exhibit variability in their luminosities over time, so single ``snapshot'' surveys of nearby galaxies may not capture the full population of HMXBs. However, even in the absence of active accretion, the donor stars of these HMXBs will themselves be faint X-ray sources that should be present in a stacked image (e.g., Section~\ref{section:stacking}. Examining the spatial clustering of YSC and bright HMXBs should therefore provide a weaker clustering signal than one would retrieve from the full population of faint and bright HMXBs, and as such we expect the clustering signal we find in Section~\ref{section:spatial}) is a lower limit on the ``true'' HMXB-YSC clustering signal.

The five galaxies that we consider here have experienced different recent SFHs, and the population of compact objects powering HMXBs may differ across galaxies. The HMXB population in the SMC is almost entirely powered by NSs with B- or Be-type companions, with no confirmed BH-HMXBs \citep{Antoniou+19,Haberl+16}. This has been explained by the SFH of the SMC, which shows a burst of star formation occurring $\sim$40 Myr ago \citep[a typical main sequence lifetime of a B-star;][]{Antoniou+19,Antoniou+10}: the most massive binaries -- those most likely to be powered by BHs -- have evolved beyond the HMXB stage, and NS-HMXBs with B-type companions are currently at maximum production. By contrast, \citet{Rangelov+11} posit that the HMXB population in NGC~4449 includes a large number of BH-HMXBs, which is consistent with the burst in star formation that occurred $\sim$10 Myr ago in that galaxy \citep{Sacchi+18}. Both the SMC and NGC~4449 show high degrees of spatial clustering between HMXBs and YSCs, but the dominant population of compact object powering the HMXBs is likely different. The colliding Antennae galaxies (Arp 244) are another example of a galaxy undergoing an extreme starburst \citep{Seille+22} in the last $\sim$10 Myr where numerous HMXBs are observed to be spatially coincident with YSCs \citep{Rangelov+12,Poutanen+13}. We show that the spatial clustering of HMXBs and YSCs is correlated with sSFR, and we suggest this correlation may be driven by differences in SFHs of the host galaxy. Such a scenario would predict the HMXBs and ultraluminous X-ray sources observed in the Antennae are dominated by extremely young BH-HMXBs.

The next most-clustered galaxy in our sample is M33, which underwent two recent periods of enhanced star formation: one $\sim$10 Myr ago, and the other $\sim40-60$ Myr ago \citep{Lazzarini+21,Garofali+18}, although the peak SFRs during these epochs were likely not as high as in the SMC or NGC~4449. The HMXB population in M33 may therefore be more mixed in terms of compact object type, yielding spatial clustering between HMXBs and YSCs that is significant but not as dramatic as observed in NGC~4449 and the SMC. The two largest galaxies in this sample -- M31 and the Milky Way -- may represent the HMXB-YSC clustering that occurs for roughly constant SFHs, at least on timescales relevant for HMXB formation and evolution.

Natal kicks during compact object formation can additionally explain both the observed displacement between HMXBs and YSCs and the apparent lack of quiescent HMXBs residing in YSCs. Observations of Galactic Be-XRBs and isolated NSs suggest that the natal kick speed distribution is bimodal, with $\sim$20-30\% of NSs receiving weak kick speeds of $45^{+25}_{-15}$ km s$^{-1}$ and the remaining experiencing stronger kicks of $\sim300$ km s$^{-1}$ or faster \citep{Igoshev20,Igoshev+21}. Kick speeds larger than $\sim100$ km s$^{-1}$ are more likely to unbind the progenitor binary and may be responsible for the population of observed isolated NSs and pulsars; NSs in HMXBs are thus more likely to have experienced weaker kicks during compact object formation \citep{Giacobbo+20}. After $\sim$10 Myr, an initial kick speed of 45 km s$^{-1}$ results in a total travel distance of $\sim$460 pc, well in line with the observed minimum separation distances between YSCs and HMXBs shown in Figure~\ref{fig:nearest_neighbor}. A high degree of clustering could either be set by a population of very recently formed NS-HMXBs receiving weak kicks (as in the SMC), or a population of BHs which received weak kicks (i.e., with similar average kick speeds as the weakly-kicked NSs) or are direct collapse sources with zero kick (as proposed for NGC~4449). In a mixed population of young BH-HMXBs and older NS-HMXBs, one would expect BH-HMXBs to be found systematically closer to their birth YSCs than the NS-HMXBs, which have had time to travel further from their birthplaces (and dilute the spatial clustering signal, as may be happening in M33).

\section*{Acknowledgements}
We thank the anonymous referee for helpful suggestions that improved the clarity of this paper. B.A.B acknowledges support from the National Aeronautics and Space Administration through \chandra\ Award Number GO0-21031X, issued by the \chandra\ X-ray Observatory Center, which is operated by the Smithsonian Astrophysical Observatory for and on behalf of the National Aeronautics Space Administration under contract NAS8-03060, as well as support from the National Science Foundation Launching Early-Career Academic Pathways in the Mathematical and Physical Sciences (LEAPS-MPS) award \#2213230. This research has made use of data obtained from the \chandra\ Data Archive and software provided by the \chandra\ X-ray Center (CXC) in the application package CIAO. This research made use of Astropy, a community-developed core Python package for Astronomy \citep{astropy:2013, astropy:2018}.

\section*{Data Availability}
All data used in this work was taken from publicly available databases and/or the cited publications.
 


\bibliographystyle{mnras}
\bibliography{references}




\bsp	
\label{lastpage}
\end{document}